\documentclass[twocolumn,superscriptaddress,prl,showpacs]{revtex4}
\usepackage{graphicx}
\usepackage{subfigure}
\usepackage{dcolumn}\usepackage{bm}
\usepackage[usenames]{color}

\def\lsim{\mathrel{\rlap{\lower4pt\hbox{\hskip1pt$\sim$}}
   \raise1pt\hbox{$<$}}}
\def\gsim{\mathrel{\rlap{\lower4pt\hbox{\hskip1pt$\sim$}}
   \raise1pt\hbox{$>$}}}

\begin{document}
\topmargin 0.0001cm
\title{First measurement  of direct  $f_0(980)$ photoproduction on the proton\\}

%
%
%
%


\newcommand*{\INFNGE}{Istituto Nazionale di Fisica Nucleare, Sezione di Genova, 16146 Genova, Italy}
\affiliation{\INFNGE}
\newcommand*{\INDIANA} {Physics Department and Nuclear Theory Center \\ Indiana University, Bloomington, Indiana 47405}
\affiliation{\INDIANA}

\newcommand*{\ANL}{Argonne National Laboratory, Argonne, Illinois 60439}
\affiliation{\ANL}
\newcommand*{\ASU}{Arizona State University, Tempe, Arizona 85287-1504}
\affiliation{\ASU}
\newcommand*{\UCLA}{University of California at Los Angeles, Los Angeles, California  90095-1547}
\affiliation{\UCLA}
\newcommand*{\CSU}{California State University, Dominguez Hills, Carson, CA 90747}
\affiliation{\CSU}
\newcommand*{\CMU}{Carnegie Mellon University, Pittsburgh, Pennsylvania 15213}
\affiliation{\CMU}
\newcommand*{\CUA}{Catholic University of America, Washington, D.C. 20064}
\affiliation{\CUA}
\newcommand*{\SACLAY}{CEA-Saclay, Service de Physique Nucl\'eaire, 91191 Gif-sur-Yvette, France}
\affiliation{\SACLAY}
\newcommand*{\CNU}{Christopher Newport University, Newport News, Virginia 23606}
\affiliation{\CNU}
\newcommand*{\UCONN}{University of Connecticut, Storrs, Connecticut 06269}
\affiliation{\UCONN}
\newcommand*{\ECOSSEE}{Edinburgh University, Edinburgh EH9 3JZ, United Kingdom}
\affiliation{\ECOSSEE}
\newcommand*{\FU}{Fairfield University, Fairfield CT 06824}
\affiliation{\FU}
\newcommand*{\FIU}{Florida International University, Miami, Florida 33199}
\affiliation{\FIU}
\newcommand*{\FSU}{Florida State University, Tallahassee, Florida 32306}
\affiliation{\FSU}
\newcommand*{\GWU}{The George Washington University, Washington, DC 20052}
\affiliation{\GWU}
\newcommand*{\ECOSSEG}{University of Glasgow, Glasgow G12 8QQ, United Kingdom}
\affiliation{\ECOSSEG}
\newcommand*{\ISU}{Idaho State University, Pocatello, Idaho 83209}
\affiliation{\ISU}
\newcommand*{\INFNFR}{INFN, Laboratori Nazionali di Frascati, 00044 Frascati, Italy}
\affiliation{\INFNFR}
\newcommand*{\ORSAY}{Institut de Physique Nucleaire ORSAY, Orsay, France}
\affiliation{\ORSAY}
\newcommand*{\ITEP}{Institute of Theoretical and Experimental Physics, Moscow, 117259, Russia}
\affiliation{\ITEP}
\newcommand*{\IHEP}{Institute for High Energy Physics, Protvino, 142281, Russia}
\affiliation{\IHEP}
\newcommand*{\JMU}{James Madison University, Harrisonburg, Virginia 22807}
\affiliation{\JMU}
\newcommand*{\UK}{University of Kentucky, Lexington, Kentucky 40506}
\affiliation{\UK}
\newcommand*{\KHARKOV}{Kharkov Institute of Physics and Technology, Kharkov 61108, Ukraine}
\affiliation{\KHARKOV}
\newcommand*{\KYUNGPOOK}{Kyungpook National University, Daegu 702-701, Republic of Korea}
\affiliation{\KYUNGPOOK}
\newcommand*{\UMASS}{University of Massachusetts, Amherst, Massachusetts  01003}
\affiliation{\UMASS}
\newcommand*{\NINP}{Henryk Niewodniczanski Institute of Nuclear Physics PAN, 31-342 Krak'ow, Poland}
\affiliation{\NINP}
\newcommand*{\UNH}{University of New Hampshire, Durham, New Hampshire 03824-3568}
\affiliation{\UNH}
\newcommand*{\NSU}{Norfolk State University, Norfolk, Virginia 23504}
\affiliation{\NSU}
\newcommand*{\UNCW}{University of North Carolina, Wilmington, North Carolina 28403}
\affiliation{\UNCW}
\newcommand*{\UAT}{North Carolina Agricultural and Technical State University, Greensboro, North Carolina 27455}
\affiliation{\UAT}
\newcommand*{\OHIOU}{Ohio University, Athens, Ohio  45701}
\affiliation{\OHIOU}
\newcommand*{\ODU}{Old Dominion University, Norfolk, Virginia 23529}
\affiliation{\ODU}
\newcommand*{\RPI}{Rensselaer Polytechnic Institute, Troy, New York 12180-3590}
\affiliation{\RPI}
\newcommand*{\RICE}{Rice University, Houston, Texas 77005-1892}
\affiliation{\RICE}
\newcommand*{\URICH}{University of Richmond, Richmond, Virginia 23173}
\affiliation{\URICH}
\newcommand*{\RIKEN}{The Institute of Physical and Chemical Research, RIKEN, Wako, Saitama 351-0198, Japan}
\affiliation{\RIKEN}
\newcommand*{\MOSCOW}{Skobeltsyn Nuclear Physics Institute, Skobeltsyn Nuclear Physics Institute, 119899 Moscow, Russia}
\affiliation{\MOSCOW}
\newcommand*{\SCAROLINA}{University of South Carolina, Columbia, South Carolina 29208}
\affiliation{\SCAROLINA}
\newcommand*{\JLAB}{Thomas Jefferson National Accelerator Facility, Newport News, Virginia 23606}
\affiliation{\JLAB}
\newcommand*{\UNIONC}{Union College, Schenectady, NY 12308}
\affiliation{\UNIONC}
\newcommand*{\UTFSM}{Universidad T\'ecnica Federico Santa Mar\'ia, Valpara\'iso, Chile}
\affiliation{\UTFSM}
\newcommand*{\VT}{Virginia Polytechnic Institute and State University, Blacksburg, Virginia   24061-0435}
\affiliation{\VT}
\newcommand*{\VIRGINIA}{University of Virginia, Charlottesville, Virginia 22901}
\affiliation{\VIRGINIA}
\newcommand*{\WM}{College of William and Mary, Williamsburg, Virginia 23187-8795}
\affiliation{\WM}
\newcommand*{\YEREVAN}{Yerevan Physics Institute, 375036 Yerevan, Armenia}
\affiliation{\YEREVAN}
\newcommand*{\NOWGWU}{The George Washington University, Washington, DC 20052}
\newcommand*{\NOWCNU}{Christopher Newport University, Newport News, Virginia 23606}
\newcommand*{\NOWCUA}{Catholic University of America, Washington, D.C. 20064}
\newcommand*{\NOWECOSSEE}{Edinburgh University, Edinburgh EH9 3JZ, United Kingdom}
\newcommand*{\NOWLANL}{Los Alamos National Laborotory, New Mexico, NM}
\newcommand*{\NOWJLAB}{Thomas Jefferson National Accelerator Facility, Newport News, Virginia 23606}

\author {M.~Battaglieri} 
\affiliation{\INFNGE}
\author {R.~De~Vita} 
\affiliation{\INFNGE}
\author {A.~P. Szczepaniak}
\affiliation{\INDIANA}

\author {K. P. Adhikari} 
\affiliation{\ODU}
\author {M.~Aghasyan} 
\affiliation{\INFNFR}
\author {M.J.~Amaryan} 
\affiliation{\ODU}
\author {P.~Ambrozewicz} 
\affiliation{\FIU}
\author {M.~Anghinolfi} 
\affiliation{\INFNGE}
\author {G.~Asryan} 
\affiliation{\YEREVAN}
\author {H.~Avakian} 
\affiliation{\JLAB}
% ...
\collaboration{The CLAS Collaboration}
    \noaffiliation
%

%
%

\date{\today}

\begin{abstract}
We report on the results of the first  measurement  
of   exclusive  $f_0(980)$ meson  photoproduction 
on protons  for  $E_\gamma=3.0 - 3.8$ GeV and 
$-t = 0.4-1.0$ GeV$^2$. Data were collected with the 
CLAS detector at the Thomas Jefferson National Accelerator Facility.
The resonance was detected via its decay in the $\pi^+ \pi^-$ channel by
performing a  partial wave analysis of the reaction $\gamma p \to  p \pi^+ \pi^-$.
Clear evidence of the $f_0(980)$ meson was found in the interference between 
$P$ and $S$ waves at $M_{\pi^+ \pi^-}\sim 1$ GeV. 
The $S$-wave  differential cross section integrated  in the mass range of the $f_0(980)$
was found to be a factor of 50 smaller than the cross section for the $\rho$ meson.
This is the first time the $f_0(980)$ meson has been measured in a photoproduction experiment.
\end{abstract}

\maketitle
\narrowtext

For a long time
most of our knowledge on  the light quark meson spectrum was obtained from hadron-induced reactions, where typically $\pi$, 
$K$, $p$ or $\bar p $ beams were used, while very
few studies with electromagnetic probes were attempted. 
Recently, high-intensity and high-quality tagged-photon beams, as the one available at JLab, have opened
a new window into this field. 
On one hand, through vector meson dominance, the photon can be effectively described as a 
virtual vector meson. On the other hand, quark-hadron duality and the point-like-nature 
of the photon coupling make it possible to describe photo-hadron interactions at the QCD level.

Spectroscopy of  low-lying scalar mesons is of particular interest. 
Recent advances in application of chiral effective field theory with dispersion relations ~\cite{Colangelo:2001df,Pelaez:2004vs,Pelaez:2003dy,Pennington:2006dg,Oller:1998hw} led to extensive investigation of this topic.
Experimental and theoretical evidence indicates
that light scalar mesons make a full $SU(3)$ flavor nonet. However, the mass spectrum ordering of the
$\sigma$, $\kappa$, $f_0(980)$, and $a_0(980)$ mesons disfavors  the naive $q \bar q$ picture.
The most natural explanation for this multiplet with an inverted mass spectrum is that these mesons are diquark-antidiquark
bound states with correct mass ordering~\cite{Jaffe,Maiani, Hooft}.
The dependence of the cross section on 
the momentum transfer $t$ and resonance   mass, which reflects  the properties of the production process,
  might shed light on the peculiar structure of these mesons.
For example, the authors of Ref.~\cite{watson} suggest that  a compact   $q{\bar q}$ system is expected to be observed
 as  a peak in the invariant mass distribution 
of the resonance decay products, while a diffuse state, {\it e.g.} a meson molecule, 
would more likely appear as a dip. 
Furthermore, the knowledge of the photoproduction cross section and spin density matrix elements is relevant
for  CP and CPT violation  studies  via $K{\bar K}$ interferometry~\cite{Isgur:2001yz}.

So far, scalar mesons have been observed in hadron-hadron collisions, $\gamma\gamma$ collisions and in 
decays of various mesons such as  $\phi, J/\Psi, D$ and $B$. Their cross sections are relatively small 
compared to the dominant production  of vector mesons; however $S$-wave parameters can be extracted 
by performing a partial wave analysis and exploiting the interference with the dominant $P$-waves.
The dominant decay mode for most of the light scalar mesons is the $\pi \pi$ channel.
Up to now the most comprehensive analyses of $\pi^+\pi^-$ photoproduction at  few GeV energies 
were performed at DESY \cite{ABBHHM,Struczinski}, SLAC~\cite{Ballam_1,Ballam_2} and Jefferson Lab~\cite{rho-clas}.
These measurements showed the dominance of  the $\rho$ resonance. 
In the analysis of the SLAC data, the angular 
dependence was parametrized in terms of $P$-wave alone, and no attempt was made 
to extract $S$-wave or higher partial waves. More recently, the HERMES 
Collaboration investigated the interference of the $P$-wave in $\pi^+\pi^-$ electroproduction 
(with $Q^2 > 3 \mbox{ GeV}^2$) with the $S$- and $D$-waves~\cite{HERMES}. 

In this work we focus on  $\pi^+\pi^-$ photoproduction 
at photon energies between
$3.0 \mbox{ GeV}$  and $3.8 \mbox{ GeV}$  in the range of momentum transfer squared 
$- t$  between $0.4 \mbox{ GeV}^2$ and $1.0 \mbox{  GeV}^2$  and present the first
analysis of the $S$-wave  photoproduction of pion pairs in the region of the $f_0(980)$. 

The present measurement was performed using the CLAS detector (CEBAF Large Acceptance Spectrometer)~\cite{B00}
at Jefferson Lab in experimental Hall B
with a bremsstrahlung photon beam produced by a primary continuous electron beam of energy $E_0$ = 4.0 GeV
hitting a gold foil  of   $10^{-4}$ radiation lengths.
A bremsstrahlung tagging system
with an
energy resolution of 0.1$\%$ $E_0$
was used to tag photons in the energy range 3.0-3.8 GeV. 
The target consisted of a 40-cm-long cylindrical cell containing liquid hydrogen at 20.4 K.
The high-intensity photon flux ($\sim  10^7 \gamma$/s) was continuously monitored during  data taking by an $e^+ e^-$
pair spectrometer located downstream of the target.
The systematic uncertainty of the photon flux has been estimated to be 10$\%$.

Outgoing hadrons were detected and identified in CLAS.
Charged particle trajectories were bent by  a toroidal magnetic 
field ($\sim 0.5$ T), which is generated by six superconducting coils.
Momentum information was obtained via tracking
through three regions of multi-wire drift chambers.
The CLAS momentum resolution for charged particles is approximately 0.5-1\% ({\bf $\sigma$}) depending on
the kinematics. The detector geometrical acceptance for each positive particle in the 
relevant kinematic region is about 40\%.
Time-of-flight scintillators (TOF) were used for hadron identification.

The interaction time of the incoming photon in the target
was measured by detecting the outgoing particles in the Start Counter (ST)~\cite{ST}.
It  consists of a set of 24 2.2-mm thick plastic scintillators 
surrounding the hydrogen cell.
Coincidences between the photon tagger and two charged particles in the CLAS detector triggered the recording of the events. 
An integrated tagged luminosity of  $\sim$70 pb$^{-1}$  was accumulated in 50 days of running. 
In total $\sim$20 TB of data were collected. 

The exclusive reaction $\gamma p \to p f_0(980)$ was  measured via 
the most sizable $f_0(980)$ decay mode
($f_0(980) \to \pi^+ \pi^- $ with $\Gamma(\pi \pi)/[\Gamma(\pi \pi)+\Gamma(K \bar K)] \sim 75 \%$~\cite{PDG}).
The final state was selected 
requiring  detection of both the proton and the $\pi^+$ in CLAS and reconstructing the $\pi^-$ 
using the missing-mass technique. About 40M events were identified after all selection cuts.
Calibrations of all detector components were performed, achieving a precision of a
few MeV in the invariant di-pion mass determination. An  experimental
resolution of a similar magnitude was evaluated
 from  Monte Carlo simulation.

The data analysis consisted of two main steps:
1) extraction of moments $\langle Y_{LM} \rangle$ of the di-pion angular distributions; 2)
fit of the  moments with a parametrization of the partial waves. In the following, we briefly outline 
the procedure, referring to a more  comprehensive  paper~\cite{PRC_2pi} for analysis  details.

\begin{figure}
\vspace{11.cm} 
\includegraphics{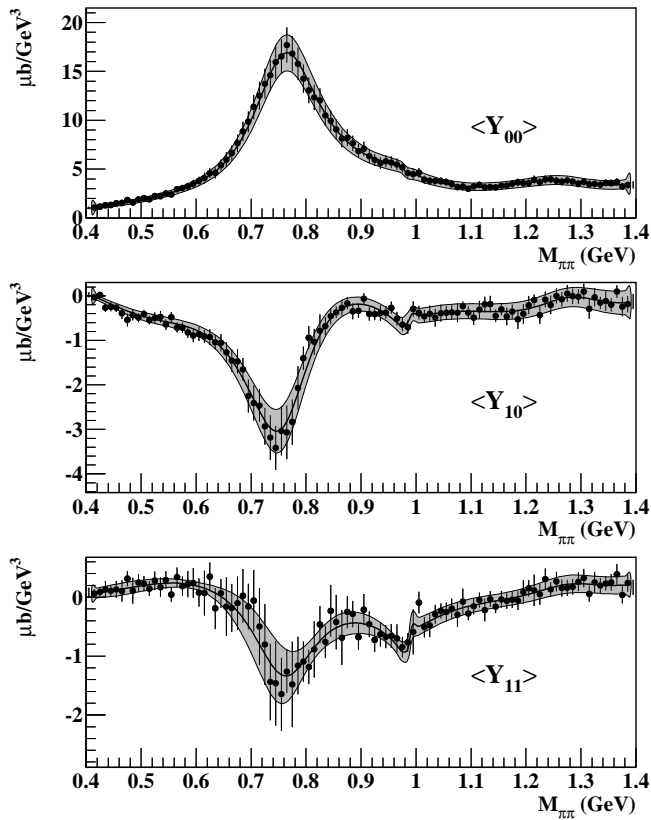}
\caption[]{Angular moments $\langle Y_{00} \rangle$ (top), $\langle Y_{10} \rangle$ (middle) and $\langle Y_{11} \rangle$ (bottom) in the 
photon energy bin $3.4 <E_\gamma< 3.6$ GeV and  momentum transfer $0.5<-t<0.6$ GeV$^2$.
Error bars include the systematic uncertainty related to the photon flux normalization and the 
moment extraction procedure. The gray band shows the result of the fit of the moments in terms of partial wave amplitudes.}
\label{moments}
\end{figure}

Moments $Y_{LM}(\Omega_\pi)$ are defined as the projection of the production cross section on spherical harmonics with defined angular momentum
$L$ and $z$-component $M$: 
\begin{equation}\label{eq:mom}
\langle Y_{LM} \rangle(E_\gamma,t,M_{\pi\pi}) = {\sqrt{4\pi}} \int d\Omega_\pi  Y_{LM}(\Omega_\pi) {{d\sigma} \over {dt dM_{\pi\pi} d\Omega_\pi}}, 
\end{equation} 
where $E_\gamma$ is the photon energy, $t$ the invariant momentum transfer to the di-pion system squared, and $M_{\pi\pi}$ its mass.
The decay angles  
$\Omega_\pi = (\theta_\pi , \phi_\pi)$ are the polar and azimuthal angles of the $\pi^ +$
in the helicity rest frame. 

The extraction of the moments from data requires that the measured angular distributions are corrected by acceptance.
The CLAS acceptance and reconstruction efficiency  were evaluated with  Monte Carlo simulations.
Events were generated  according to three-particle phase space in the same photon energy range as the experiment,
processed by a GEANT-based code that included knowledge of the detector geometry and response to traversing particles,
and reconstructed using the same analysis procedure that was applied to the data.
Moments were expanded in a model-independent way in two  sets of basis functions 
and, after weighting with Monte Carlo simulations,
they were fitted to the data by  maximizing a likelihood function.
This was built on an event-by-event basis, to avoid  binning of the experimental data.
In the first case, the parametrization was given 
in terms of 
$amplitudes$, while in the second, $moments$ were directly used~\cite{Grayer}.
In both cases the number of basis functions was limited for practical reasons.
In the kinematic range $3.0<E_\gamma<3.8$ GeV, $0.4<-t<1.0$ GeV$^2$ and $0.4 < M_{\pi\pi} <1.4$ GeV,
moments with $L \leq 4$ and $|M| \leq L$ 
were calculated as an average of results obtained with the two parametrizations.

Detailed systematic studies  were performed using  both Monte Carlo simulations 
and real data to ensure the validity of the approximations and to study possible 
effects related to the basis truncation and the detector acceptance. 
The comparison of results obtained by the different methods was used to estimate the systematic uncertainty
related to the analysis  procedure. We found that the variation in the moments obtained from the different procedures
is larger than 
the statistical 
 uncertainty and larger than other sources of systematic uncertainty, such as event selection cuts, detector resolution and inefficiency.
The final uncertainty  was then obtained by summing in quadrature the fit uncertainty given by MINUIT, 
the uncertainty associated with the photon flux determination,
and the above-mentioned uncertainty on the moment extraction procedure.

The plots in Fig.~\ref{moments} show the moments $\langle Y_{00} \rangle$, $\langle Y_{10} \rangle$ 
and $\langle Y_{11} \rangle$ in a selected  $E_\gamma$ and  $t$ bin. From Eq.~\ref{eq:mom} it is 
straightforward to show that 
moment $\langle Y_{00} \rangle$ corresponds to the differential cross section $d\sigma / dt dM_{\pi\pi}$.
As expected this is dominated by the contribution of the $\rho$ meson in the $P$-wave 
shown by the prominent peak at $M_{\pi\pi} \sim 0.77\mbox{ GeV}$.
In moments $\langle Y_{10} \rangle$ and $\langle Y_{11} \rangle$, the contribution of the $S$-wave is 
maximum and enters via interference with the dominant $P$-wave.

The second step of the analysis consisted of extracting the partial wave amplitudes from the angular moments.
These can be  expressed as bi-linear in terms of the amplitudes 
$a_{lm}=a_{lm}(\lambda,\lambda',\lambda_\gamma,E_\gamma,t,M_{\pi\pi})$ with  angular momentum $l$ and $z$-projection $m$
(in the chosen reference system $m$ coincides with the helicity of the di-pion system):
\begin{equation}
\langle Y_{LM} \rangle \propto   
\sum_{l'm',lm , \lambda, \lambda'} C(l'm',lm, LM) \times  a_{lm} \;  a^*_{l'm'}\label{Ytheor}, 
\end{equation}
where $\lambda$ and  $\lambda'$ are the initial and final nucleon 
helicity, respectively, $\lambda_\gamma$ is the helicity of the photon, and $C$ are  Clebsch-Gordan  coefficients. 
Each amplitude was expressed as a linear combination  of $\pi\pi$ 
amplitudes of fixed isospin, $a_{lm,I}$ with $I=0,1,2$.
The number of waves  was  reduced restricting the analysis to  $|m| \le 1$, since 
$m=2$ waves 
are expected to be small in the mass range under investigation~\cite{PRC_2pi}.
The photon helicity was  restricted to  $\lambda_\gamma  = +1$
since the other amplitudes are related by parity conservation.
As a consequence only  three values of  $m$ have to be considered: 
$m = +1$, which corresponds to a non-helicity flip ($s$-channel helicity conserving) amplitude, 
expected to be  dominant, and $m = 0, -1$ that correspond 
to one and two units of helicity flip, respectively. 
In the case of the $S$-wave ($l=m=0$), only one amplitude is considered. 
The dependence on the nucleon helicity was simplified as follows. 
For a given $l,m,E_\gamma,t$ set,  there are   four independent  
partial wave amplitudes corresponding to the four combinations of initial and final nucleon helicity.
In general it is expected that dominant amplitudes require no helicity flip~\cite{Ballam_2}.
On the other hand, we found that  at least two amplitudes were necessary to reproduce the data:
therefore for each $l,m$, with $|m|\le 1$ we used two sets of amplitudes corresponding
to  helicity non-flip and helicity-flip of one unit.

For each helicity  state  of the target $\lambda$, recoil nucleon $\lambda'$, and  $\pi\pi$ system $m$,  in a given $E_\gamma$ and $t$ bin, 
the corresponding  helicity amplitude  $a_{lm}(s=M^2_{\pi\pi})$, was expressed using a dispersion relation~\cite{Aitchison:1976nk,Bowler:1975my,Basdevant:1976jg} as follows:
\begin{eqnarray}
\label{disp}
a_{lm,I}(s) & = & {1\over 2} [ I  + S_{lm,I}(s) ] \tilde a_{lm,I}(s)\\ \nonumber
& - &  {  1\over {\pi}} D_{lm,I}^{-1}(s) PV  \int_{s_{th}} ds' {{ N_{lm,I}(s') \rho(s')  \tilde a_{lm,I}(s')} \over {s'-s}}, 
\end{eqnarray} 
where $PV$ represents the principal value of the integral and $\rho$ corresponds to  the phase space term.
In this expression, 
$N_{lm,I}$ and $D_{lm,I}$ can be written in terms of the scattering matrix of $\pi\pi$ scattering, 
chosen to reproduce the known phase  shifts, inelasticities~\cite{PDG,Oller:1998hw},
and the  isoscalar ($l=S,D$), isovector ($l=P,F$) 
and isotensor ($l=S,D$) amplitudes in the range $0.4 \mbox{ GeV} < \sqrt{s}  < 1.4 \mbox{ GeV}$.
$I$ and $S_{lm,I}$ are matrices in the multi-channel space ($\pi\pi$, $KK$) relevant for the mass range considered in this analysis. 
The amplitude $\tilde a_{lm,I}$
represents our ignorance about the production process. 
Since discontinuities are taken into account by functions $N_{lm,I}$ and $D_{lm,I}$, $\tilde a_{lm,I}(s)$ does not have singularities
for $s > 4m_\pi^2$ and can be expanded in a polynomial function. This was chosen to be of second order and its coefficients
are the partial wave analysis parameters that were extracted by the simultaneous fit of the angular moments defined in  Eq.~\ref{Ytheor}.
All amplitudes but the  scalar-isoscalar are saturated by the $\pi\pi$ state. For the scalar-isoscalar amplitude, the $K\bar K$ 
channel was also included. In addition, to reduce sensitivity to the large energy  behavior of the  ($\pi\pi$,$K\bar K$)  amplitudes,
 the real part of the integral was subtracted and replaced by a polynomial in $s$, whose coefficients   were also fitted.
The imaginary part of the integral in Eq.~\ref{disp}  represents the production of long-lived (on-shell) meson pairs
corresponding to  the non-resonant part of the scattering process.
The real part of the same integral represents the direct resonant production that, in the absence of the on-shell part, 
would lead to the typical Breit-Wigner shape.

Partial waves $a_{lm}$ up to  $l=3$ ($F$ wave) 
were determined fitting  all moments $\langle Y_{LM}\rangle$ with $L \leq 4$ and $|M|\leq min(L,2)$.
\begin{figure} 
\vspace{8.cm} 
\includegraphics{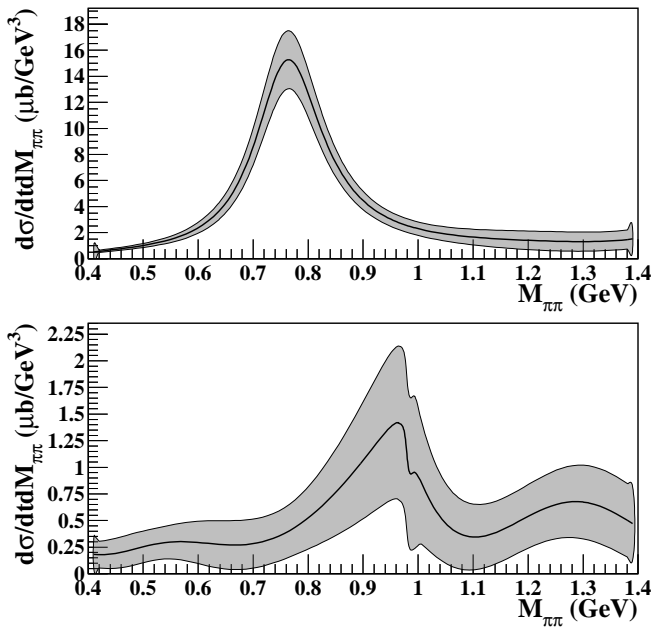}
\caption[]{ 
Partial wave  cross sections in the same kinematic bin as Fig.~\ref{moments}. The top and bottom panels show the 
$P$- and  the $S$-wave, respectively. The width of the bands represents the uncertainty estimated as the sum in quadrature of
statistical and systematic uncertainties as explained in the text. }
\label{S-wave}
\end{figure}
Results of the fit are shown as a gray band in Fig.~\ref{moments}  on top of the experimental angular moments 
$\langle Y_{00} \rangle$, $\langle Y_{10} \rangle$ and $\langle Y_{11} \rangle$ in a selected  $E_\gamma$ and  $t$ bin.
As stated above, the contribution of the $S$-wave is maximum in moments 
$\langle Y_{10} \rangle$ and $\langle Y_{11} \rangle$.
In particular the large structure at the  $\rho$ mass in $\langle Y_{11} \rangle$ is due to 
the interference of the $S$ wave with the dominant, helicity-non-flip wave, $P_{m=1}$  ($\lambda_\gamma=1 \to m=1$).
In moment $\langle Y_{10} \rangle$ the same structure is due to the
interference with the $P_{m=0}$ wave, corresponding to  one unit of helicity flip ($\lambda_\gamma=1 \to m=0$).
A second dip near $M_{\pi\pi} = 1\mbox{ GeV}$ is clearly visible and  corresponds to the 
direct production of a resonance that we interpret as the $f_0(980)$. The mass and  width of this
structure are compatible with the PDG values ($M = 980 \pm 10$ MeV and $\Gamma = 40-100$ MeV ~\cite{PDG}).

It should be noted that moments of the $\pi^+\pi^-$ angular distribution can be affected
by baryon resonances decaying to $\pi^+p$ and $\pi^- p$.
These contributions represent a background for our analysis 
but,  having a smooth dependence on the  di-pion mass, they cannot create narrow
structures in these observables. In addition, they 
are expected to be small for low moments and limited values of $M_{\pi\pi}$ ($\lsim 1.1$ GeV) that are the focus of this analysis.

The $P$  and  $S$ partial wave differential cross sections $d\sigma/dtdM_{\pi\pi}$ are shown in  Fig.~\ref{S-wave}.
As expected, the  $S$-wave photoproduction is suppressed compared to  the $P$-wave, which is dominated by the $\rho$ meson.
This can be explained within Regge theory because vector meson production can proceed via  Pomeron exchange,  
while scalars require exchange of  reggeons 
that become suppressed as energy increases.  
As a test of the whole procedure, the differential cross section $d\sigma/dt$ for the reaction 
$\gamma p \to p \rho \to p \pi^+ \pi^-$ was derived  integrating  the $M_{\pi\pi}$ mass from 0.4 GeV to 1.2 GeV.
Comparison with previous CLAS measurements~\cite{rho-clas} and ABBHHM Collaboration data~\cite{ABBHHM}
shows good agreement,  giving confidence in the partial wave analysis. More details can be found in Ref.~\cite{PRC_2pi}.

\begin{figure}
\vspace{6.cm} 
\includegraphics{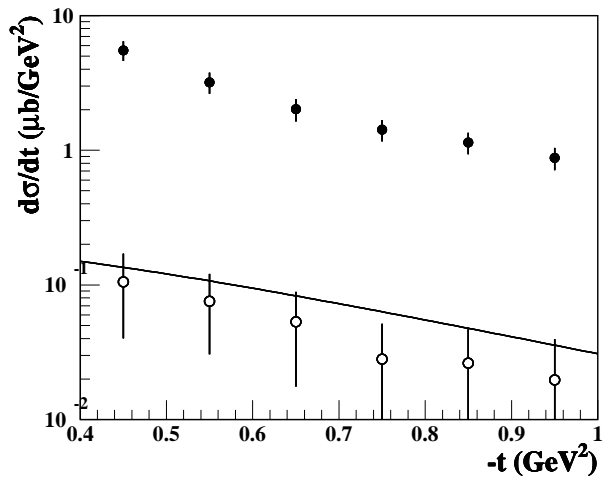}
\caption[]{Partial wave differential cross sections  $d\sigma/dt$ in the photon energy range $E_\gamma=3.0 - 3.8$ GeV,
for the $P$-wave (solid dots) 
and $S$-wave (open circles) integrated in the 
$M_{\pi\pi}$ mass range 0.4-1.2 GeV  and  $0.98\pm0.04$ GeV, respectively.
The error bars include statistical and systematic uncertainties summed in quadrature.
The line is a model prediction for the $S$-wave from Refs.~\cite{Bibrzycki:2003mq,Ji:1997fb}.}
\label{dsdt-S-wave}
\end{figure}

The $S$-wave shows a clear variation in the vicinity of the $f_0(980)$.
However, the resonance component seems to be embedded in a coherent background. 
The evidence of the  $f_0(980)$ signal in the $S$-wave is a sign that photoproduction may 
indeed be a good tool for accessing meson resonances other than vector meson states~\cite{halld}. 
The total $S$-wave differential cross section $d\sigma/dt$ in the region of the $f_0(980)$ was obtained integrating 
  the  $M_{\pi\pi}$ mass in the range $0.98\pm0.04$ GeV.
Differential cross sections $d\sigma/dt$  in $E_\gamma =  3.4 \pm 0.4$ GeV, 
for $P$-wave (solid dots)  and $S$-wave (open circles) obtained as described above
are shown  in Fig.~\ref{dsdt-S-wave}. 
The solid line is a   prediction for the $S$-wave of a model based on Regge exchanges~\cite{Bibrzycki:2003mq,Ji:1997fb}.
This  was normalized to DESY  $K^+K^-$ photoproduction data~\cite{Behrend} and was able to reproduce 
 the $S$-wave measured in the same channel at Daresbury~\cite{Barber}.
The agreement of the calculation with our data suggests that
the $\pi^+\pi^-$ $S$-wave cross section extracted here is consistent with the measurement in the   $K^+K^-$ channel.
It also indicates that the present  data  can be used in  phenomenological analyses
that, exploiting the  point-like nature of photon interactions, will  provide information 
about  the resonance structure and production mechanisms.
A detailed comparison between theory and the measured cross section will be the subject of future investigations.

In summary, we measured  $\pi^+\pi^-$ photoproduction 
in the photon energy range $E_\gamma =3.0 - 3.8 \mbox{ GeV}$ and momentum transfer range
$0.4 \mbox{ GeV}^2 < -t < 1.0 \mbox{ GeV}^2$  performing a partial wave analysis.
Moments of the di-pion angular distribution were parametrized in terms of production amplitudes,
expressed as bi-linear in the partial waves, and fitted to the experimental data.
The systematic uncertainty related to the whole procedure was estimated 
performing the analysis using different procedures and approximations.
As expected, the dominant partial wave was found to be the one associated with the 
helicity-non-flip $\rho(770)$ production. 
As a test, the $\rho$ photoproduction cross section was extracted and found to be consistent
with previous measurements.
The interference between $P$ and $S$ waves at $M_{\pi\pi} \sim 1\mbox{ GeV}$ clearly indicates 
the presence of the $f_0(980)$ resonance. 
This is the first time the $f_0(980)$ meson has been measured in a photoproduction experiment. 
Using a parametrization of the individual waves based on dispersion relations, we were able to extract the total 
 $S$-wave differential cross section
in the mass range  of this scalar meson. 
In the lowest accessible  range of the momentum transfer, $-t = 0.4-0.5 \mbox{ GeV}^2$, 
the differential cross section was found to be $d\sigma/dt = 0.11 \pm 0.06\mbox{ }\mu b/\mbox{GeV}^2$, 
which is a factor of 50 smaller than the cross section for the $P$-wave integrated in the  $\rho$ mass range.  

We would like to acknowledge the outstanding efforts of the staff of the Accelerator
and the Physics Divisions at Jefferson Lab that made this experiment possible. 
This work was supported in part by  the  Italian Istituto Nazionale di Fisica Nucleare, 
the French Centre National de la Recherche Scientifique
and Commissariat \`a l'Energie Atomique, 
the U.S. Department of Energy and National Science Foundation, 
and the Korea Science and Engineering Foundation.
Jefferson Science Associates, LLC,
operates Jefferson Lab for the United States
Department of Energy under U.S. DOE contract DE-AC05-
060R23177.

\end{document}